\documentclass[12,pre,preprint]{revtex4}

\usepackage{amsfonts}
\usepackage{amsmath}
\usepackage{amssymb}
\usepackage{theorem}

\newcommand{\one}{\mbox{\tt 1}\hspace{-0.057 in}\mbox{\tt l}}
\newcommand{\natuerl}{{\mathbb N}}

\newcommand{\Tr}{\mbox{\rm\small Tr\ }}

\newcommand{\balphas}{\vec{\mbox{$\scriptstyle \alpha$}}}

\newcommand{\bbetas}{\vec{\mbox{$\scriptstyle \beta$}}}

\begin{document}
\title{Decoherence properties of arbitrarily long
        histories\footnote{in Proceedings of the
                  7th International Conference on
				  Quantum Communication,
                  Measurement and Computing (QCMC'04),
                  edited by Stephen M. Barnett 
				  (AIP Press, Melville, NY, 2004). }}
\author{Artur Scherer\footnote{the corresponding author:
artur.scherer@uni-konstanz.de; a.scherer@rhul.ac.uk}}
\author{Andrei N.\ Soklakov\footnote{a.soklakov@rhul.ac.uk}}
\address{Department of Mathematics, Royal Holloway, University of
        London, Egham, Surrey, TW20 0EX, UK.}

\begin{abstract}
Within the decoherent histories formulation 
of quantum mechanics, we consider arbitrarily
long histories constructed from a fixed projective
partition of a finite-dimensional Hilbert space.
We review some of the decoherence properties of
such histories including simple necessary
decoherence conditions and the dependence
of decoherence on the initial state.
Here we make a first step towards generalization of
our earlier results 
[Scherer and Soklakov, e-print: quant-ph/0405080, (2004)
and Scherer et al., Phys.\ Lett.\ A {\bf 326}, 307, (2004)]
to the case of approximate decoherence. \\ 



\end{abstract}
\date{15 Sept, 2004}
\maketitle


\section{Introduction}
The formalism of decoherent histories was introduced 
to provide a self-contained description of closed quantum
systems that does not rely on either the external observer
nor on the existence of classical 
devices~\cite{Griffiths1984,Omnes1988,Gell-Mann1990,
Dowker1992,Gell-Mann1993}. It has been successfully
applied in various fields of quantum theory including
quantum cosmology~\cite{Hartle1997}, derivation of  
effective classical dynamics from the fundamental 
quantum dynamical laws ~\cite{Dowker1992,Gell-Mann1993,Halliwell1998},
and the study of the coarse-grained evolution of iterated quantum 
maps~\cite{Soklakov2002}. 
Recently the formalism of decoherent histories has 
also been applied for investigating classicality 
in quantum information processing~\cite{Poulin2002}.  

The decoherent histories formalism predicts probabilities
for quantum histories, i.e.\ ordered sequences 
of quantum-mechanical \lq\lq propositions''. 
Mathematically, these propositions are represented
by projectors. In particular, an exhaustive set of
mutually exclusive propositions corresponds to a
complete set of mutually orthogonal projectors.
Due to quantum interference, however, one cannot
always assign probabilities to a set of histories
in a consistent way. For this to be possible, 
the set of histories must be decoherent. 
Decoherence of histories ensures that
the assigned probabilities obey the standard 
probability sum rules.

In Refs.~\cite{Scherer2004A,Scherer2004B} a special,
but very important class of histories was introduced,
namely, the class of histories that are constructed
from a fixed projective partition of a finite-dimensional
Hilbert space. In the context of exact decoherence 
a number of results were obtained regarding decoherence
properties of such histories. In particular,
simple necessary decoherence conditions were
derived and the dependence of decoherence
on the initial state was investigated.
In this paper we give a brief review
of these results and make a first 
step towards generalizations of these results
to the case of {\em approximate decoherence}.

The paper is organized as follows. After introducing
our framework we first review the results obtained 
in~\cite{Scherer2004A,Scherer2004B} for the case of
exact decoherence. In the second part of the paper 
approximate decoherence of histories is introduced 
and the corresponding implications examined.

\section{Our Setting}

\noindent {\bf Definition 1:}
A set of projectors $\{P_{\mu}\}$ on a Hilbert space $\cal H$ is 
called a projective {\em partition} of $\cal H$, if $\:\forall\, \mu, 
\mu'\,:\;\:P_{\mu}P_{\mu'}=\delta_{\mu\mu'}P_{\mu}\:$ and $\:\sum_{\mu}P_{\mu}
=\one_{\cal H}$. Here, $\one_{\cal H}$ denotes the unit operator. 
We call a projective partition  {\em
  fine-grained\/} if all projectors are one-dimensional, 
i.e., if $\,\forall\,\mu\;$
$\mbox{dim}\big(\mbox{supp}(P_{\mu})\big)=1$, and {\em coarse-grained} otherwise.

\noindent{\bf Definition 2:}
Given a projective partition $\{P_{\mu}\}$ of a Hilbert space $\cal H$, 
let us denote by $\mathcal{K}[\{P_{\mu}\}\,;\,k\,]:=\big\{h_{\balphas}\,:\:
h_{\balphas}=\left(P_{\alpha_{1}},P_{\alpha_{2}},
 \dots, P_{\alpha_{k}}\right)\in\{P_{\mu}\}^k\big\}$ the 
corresponding exhaustive set of mutually exclusive histories of length
$k$. Histories are thus defined to be ordered sequences of projection
operators, corresponding to quantum-mechanical propositions. 
Note that we restrict ourselves to histories constructed from a 
{\em fixed} projective partition: the projectors $P_{\alpha_{j}}$ within the 
sequences are all chosen from the same partition for all 
\lq\lq times''  $j=1,\ldots,k$.

\noindent{\bf Definition 3:}
Given a Hilbert space $\cal H$ and a projective partition
$\{P_{\mu}\}$ of $\cal H$, we denote by $\mathcal{S}$ the 
set of all density operators on $\cal H$ and by
$\mathcal{S}_{\{P_{\mu}\}}$
the discrete set of {\em \lq\lq partition states''} induced by the partition 
$\{P_{\mu}\}$ via normalization:   
$\mathcal{S}_{\{P_{\mu}\}}:=
\big\{P_{\nu}/ \Tr[P_{\nu}]\,:\,P_{\nu}
\in\{P_{\mu}\}\big\}\,.$ 
Furthermore, a state $\rho\in\mathcal{S}$ is called 
{\em classical with respect to (w.r.t.) the partition  $\{P_{\mu}\}$}
if it is block-diagonal w.r.t.\ $\{P_\mu\}$, i.e., if 
$\,\,\rho = \sum_{\mu}P_{\mu}\,\rho\, P_{\mu}\;$. 
The set of all such classical states is denoted by 
$\mathcal{S}^{cl}_{\{P_{\mu}\}}$. \\

An initial state $\rho\in\mathcal{S}$ and a unitary  
dynamics generated by a unitary map $U:\cal H\rightarrow\cal H$ induce 
a probabilistic structure on the event algebra 
associated with $\mathcal{K}[\{P_{\mu}\}\,;\,k]$, 
if certain consistency conditions are fulfilled.
These are given in terms of properties of the {\em decoherence functional\/}
$\mathcal{D}_{U,\,\rho}\,[\cdot,\cdot]$ on 
$\mathcal{K}[\{P_{\mu}\}\,;\,k\,]\times
\mathcal{K}[\{P_{\mu}\}\,;\,k\,]\,$, defined by 
\begin{equation} 
\mathcal{D}_{U,\,\rho}\,[h_{\balphas},h_{\bbetas}]:=
\mbox{Tr}\left[C_{\balphas}\,\rho\,C_{\bbetas}^{\dagger}\right]\:,
\end{equation}
where 
\begin{eqnarray} 
C_{\balphas}& :=&\left(U^{\dagger\,k}P_{\alpha_k}U^k\right)
\left(U^{\dagger\,k-1}P_{\alpha_{k-1}}U^{k-1}\right)\dots
 \left(U^{\dagger}P_{\alpha_1}U\right)\nonumber\\
&=&U^{\dagger\,k}P_{\alpha_k}UP_{\alpha_{k-1}}U\dots
  P_{\alpha_2}UP_{\alpha_1}U \label{Classoperators}
\end{eqnarray}
The set $\mathcal{K}[\{P_{\mu}\}\,;\,k\,]$ is said to be 
{\em decoherent\/} or {\em consistent\/} with respect to a given 
unitary map $U:\cal H\rightarrow\cal H$ and a given initial 
state $\rho\in\mathcal{S}$, if 
\begin{equation}  \label{eq:consistency}
\mathcal{D}_{U,\,\rho}\,[h_{\balphas},h_{\bbetas}]
\propto \delta_{\balphas\bbetas}\equiv
\prod_{j=1}^k\delta_{\alpha_j \beta_j}
\end{equation}
for all $h_{\balphas},h_{\bbetas}\in\mathcal{K}[\{P_{\mu}\}\,;\,k\,]$.
These are the consistency conditions. 
If they are fulfilled, probabilities  
may be assigned to the histories   
and are given by the diagonal elements 
of the decoherence functional, 
$p[h_{\balphas}]=\mathcal{D}_{U,\,\rho}\,[h_{\balphas},h_{\balphas}]$.
The conditions given above are known as 
{\em medium decoherence\/}~\cite{Gell-Mann1993}.
It has recently been shown that consideration
of the weaker consistency conditions that had been proposed 
in the literature~\cite{Griffiths1984,Omnes1988} 
is problematic~\cite{Diosi2004}.

\section{Motivation}

Whether the decoherence condition~(\ref{eq:consistency}) 
is fulfilled or not depends on the initial state, 
the unitary dynamics of the system and the propositions 
from which the histories are constructed. 
The dependence on the initial state is connected to 
one of the central questions of the decoherence programme:  
the question of how the classical features of our world 
emerge from the initial quantum state of the Universe.

A rather more technical motivation for the research
presented in this paper comes from the need of
simpler decoherence conditions.  
In general, it is very difficult to decide whether 
a given set of histories is decoherent. With increasing 
length of the histories checking the decoherence
conditions~(\ref{eq:consistency}) soon becomes extremely 
cumbersome. This is especially true when the system dynamics 
is difficult to simulate as, e.g., in the case of chaotic quantum maps, 
for which typically only the first iteration is known in  closed 
analytical form. 
Establishing decoherence directly, by computing the off-diagonal 
elements of the decoherence functional, would require enormous 
computational resources in the case of large history lengths.  
It would therefore be of great practical importance 
to have a simple decoherence criterion that uses only 
a single iteration of the unitary map. 
Sufficient conditions of this type can trivially be found. 
Our analysis concentrates on {\em necessary} single-iteration 
decoherence conditions.

\section{Results for exact decoherence}

In~\cite{Scherer2004B} we have proven the following
theorem.

\noindent{\bf Theorem 1:}
{\it Let a projective partition}  $\{P_{\mu}\}$ 
{\it of a finite dimensional Hilbert space $\cal H$ and a unitary 
map} $U$ {\it on} $\cal H$ {\it be given. Then the following three 
statements are equivalent:}
\begin{eqnarray}
\label{eq:decoherence_all_P_{mu}}
&(a)&
\forall \,\rho\in
\mathcal{S}_{\{P_{\mu}\}}\;
\forall\, k\in\hspace{-0.5mm}\natuerl \;\;\forall\,h_{\balphas},
h_{\bbetas}\in\mathcal{K}[\{P_{\mu}\}\,;\,k\,]:
\;\mathcal{D}_{U,\,\rho}\,[h_{\balphas},h_{\bbetas}]
\propto  \delta_{\balphas\bbetas}\nonumber\\
\label{eq:commutativity_P}
&(b)&
\forall\,P_{\mu'}, P_{\mu''}\in\{P_{\mu}\}\;\,\forall\, n\in\natuerl
\,:\;\;\left[U^nP_{\mu'}(U^{\dagger})^n\,,\,P_{\mu''}\right]=0\nonumber\\
\label{eq:decoherence_all_rho}
&(c)&
\forall \,\rho\in\mathcal{S}\;\;
\forall\, k\in\hspace{-0.5mm}\natuerl \;\;\forall\,h_{\balphas},
h_{\bbetas}\in\mathcal{K}[\{P_{\mu}\}\,;\,k\,]:
\;\mathcal{D}_{U,\,\rho}\,[h_{\balphas},h_{\bbetas}]
\propto  \delta_{\balphas\bbetas}\;.\nonumber
\end{eqnarray}

The implication (a)$\Rightarrow $(c) of the theorem
leads to an interesting 
conclusion concerning the dependence of decoherence on the initial state: 
the decoherence of histories of arbitrary length for all initial 
states from the set ${\cal S}_{\{P_\mu\}}$ implies decoherence  
of such histories for arbitrary initial states $\rho\in\mathcal{S}$. 
Note that the set ${\cal S}_{\{P_\mu\}}$
can be viewed as the smallest natural set of states that 
is associated with our framework. It is discrete and
may consist of just two elements (in the case of \lq\lq yes-no'' propositions).
The set ${\cal S}$, on the other hand, contains the continuum
of all possible states. 

The above theorem also provides a necessary 
single-iteration decoherence condition that is applicable 
to arbitrary coarse-grainings. This generalizes  
the simple condition derived in an 
earlier paper~\cite{Scherer2004A} 
for the case of fine-grained histories. 
In~\cite{Scherer2004A} it was shown that, in the case of 
fine-grained partitions, sets of histories of arbitrary length decohere 
for all classical initial states $\rho\in\mathcal{S}^{cl}_{\{P_{\mu}\}}$ 
{\em only if} the unitary dynamics 
preserves the classicality of states, i.e.\ {\em only if} 
\begin{equation} \label{ClassicalityPreservation} 
\forall\rho\in\mathcal{S}^{cl}_{\{P_{\mu}\}}\,:\;
U\rho U^{\dagger}\in \mathcal{S}^{cl}_{\{P_{\mu}\}}.
\end{equation} 
Unfortunately, condition~(\ref{ClassicalityPreservation}) fails to
be necessary\footnote{Condition~(\ref{ClassicalityPreservation}) 
is trivially a sufficient single-iteration decoherence condition 
both in the fine-grained as well as in the 
coarse-grained case.} in the coarse-grained case~\cite{Scherer2004A}. 
The above theorem provides a necessary single-iteration condition 
that now applies to arbitrary coarse-grainings and is equivalent 
to~(\ref{ClassicalityPreservation}) in the fine-grained case.  
According to the implication (a)$\Rightarrow $(b) of the theorem the 
medium decoherence condition is satisfied for all classical initial states 
$\rho\in\mathcal{S}^{cl}_{\{P_{\mu}\}}$ and arbitrarily long histories, i.e.,
\begin{equation}  
\forall \,\rho\in\mathcal{S}^{cl}_{\{P_{\mu}\}}\;
\forall\, k\in\hspace{-0.5mm}\natuerl \;\;\forall\,h_{\balphas},
h_{\bbetas}\in\mathcal{K}[\{P_{\mu}\}\,;\,k\,]:
\;\mathcal{D}_{U,\,\rho}\,[h_{\balphas},h_{\bbetas}]
\propto  \delta_{\balphas\bbetas}\;,
\end{equation}
{\em only if the following necessary condition is fulfilled:} 
\begin{equation}  
\forall\,P_{\mu'}, P_{\mu''}\in\{P_{\mu}\}
\,:\;\;\left[UP_{\mu'}U^{\dagger}\,,\,P_{\mu''}\right]=0\,.
\label{nec_singleiteration_comm_cond}
\end{equation}

\section{Generalisation to approximate decoherence}
Condition~(\ref{eq:consistency}) is the condition for 
{\em exact} decoherence. In most of physical models, however, 
decoherence of histories can be established only {\em approximately}  
(cf.\ e.g. \cite{Dowker1992}). It is therefore desirable to generalise the 
above results to the case of approximate decoherence. Let us first explain 
what is meant by approximate decoherence. An absolutely consistent assignment of 
probabilities to a given set of histories requires that whenever we 
bundle up the given histories to coarser-grained histories then 
the probability for each such coarser-grained history must be equal 
to the sum of the probabilities for its constituent finer-grained histories, 
and this has to be true for all possible coarse-grainings. If these   
probability sum rules are fulfilled only approximately, for all 
possible coarse-grainings of a given set of finer-grained histories, 
then we get an approximately consistent assignment of probabilities 
and call the given set of histories approximately decoherent. 
Quantitatively, one requires that the probability sum rules are 
satisfied to some order $\epsilon$, meaning that the interference terms 
are suppressed by a very small factor $\epsilon\ll 1$ compared to the 
sums over the probabilities, for all possible coarse-grainings. 
A condition that proved to be useful for approximate decoherence 
is (cf.~Ref.~\cite{Dowker1992} and 
\footnote{In \cite{Dowker1992} a weaker 
condition was proposed, with
$|\,\mbox{Re}\,\mathcal{D}_{U,\,\rho}\,[h_{\balphas},h_{\bbetas}]\,|$ 
instead of $|\,\mathcal{D}_{U,\,\rho}\,[h_{\balphas},h_{\bbetas}]\,|$ 
on the left-hand side of the inequality~(\ref{approx-dec-cond_1}).})
\begin{equation}\label{approx-dec-cond_1}
\left|\,\mathcal{D}_{U,\,\rho}\,[h_{\balphas},h_{\bbetas}]\,\right| 
< \epsilon\,\left(\mathcal{D}_{U,\,\rho}\,[h_{\balphas},h_{\balphas}]
\mathcal{D}_{U,\,\rho}\,[h_{\bbetas},h_{\bbetas}]\right)^{\frac{1}{2}}\;
\;\;\,\mbox{for}\;\;
h_{\balphas}\not=h_{\bbetas}\:.
\end{equation}
In \cite{Dowker1992} it was shown that with this condition {\em most}    
(in a statistical sense) probability sum rules are satisfied to order 
$\epsilon$ provided the number of all possible histories $h_{\balphas}$    
is large. Here we assume a stronger condition, which 
guarantees that {\em all} probability sum rules are satisfied to the 
order $\epsilon$, for all possible coarse-grainings, namely,
\begin{equation}\label{approx-dec-cond_2}
\left|\,\mathcal{D}_{U,\,\rho}\,[h_{\balphas},h_{\bbetas}]\,\right| 
< \epsilon\,\frac{\left(\mathcal{D}_{U,\rho}\,[h_{\balphas},h_{\balphas}]
\mathcal{D}_{U,\,\rho}\,[h_{\bbetas},h_{\bbetas}]\right)^{\frac{1}{2}}}
{|\mathcal{K}[\{P_{\mu}\}\,;\,k\,]\,|}\;
\;\;\,\mbox{for}\;\;
h_{\balphas}\not=h_{\bbetas}\:,
\end{equation}
where $|\mathcal{K}[\{P_{\mu}\}\,;\,k\,]\,|$ denotes the number 
of elements in the set $\mathcal{K}[\{P_{\mu}\}\,;\,k\,]$, which 
is the number of all possible histories $h_{\balphas}$.  
It is bounded from above by $d^k$ with $d$ being the dimension of the 
Hilbert space, $d=\mbox{dim}\,{\cal H}$. 

The only difficult part in the proof of Theorem 1 was 
to show the implication \lq\lq(a)$\Rightarrow $(b)''. 
As a first step towards proving analogous
results for approximate decoherence, we confine ourselves
to generalizing just this part 
of Theorem~1. Instead of exact 
decoherence we now assume approximate decoherence of 
histories for  arbitrary history lengths~$k$ and for 
all initial states $\rho\in{\cal S}_{\{P_\mu\}}$, i.e., we 
replace in the statement (a) of Theorem~1 the exact decoherence 
condition~(\ref{eq:consistency}) by our approximate decoherence 
condition~(\ref{approx-dec-cond_2}). The task now
is to show that the statement~(b) of Theorem~1 
is still implied in some approximate sense. 
The derivation of this implication can only be
sketched in this paper. It is based on a Lemma
on {\it uniform recurrence in finite dimensional
Hilbert spaces}~\cite{Scherer2004A} and is done
in a similar way as in the proof of Theorem~1 in
Ref.~\cite{Scherer2004B}. Using the trivial relation 
\begin{equation}
\left(\mathcal{D}_{U,\rho}\,[h_{\balphas},h_{\balphas}]
\mathcal{D}_{U,\rho}\,[h_{\bbetas},h_{\bbetas}]\right)^{\frac{1}{2}}
\le\frac{1}{2}\left(\mathcal{D}_{U,\rho}\,[h_{\balphas},h_{\balphas}]+
\mathcal{D}_{U,\rho}\,[h_{\bbetas},h_{\bbetas}]\right)
\end{equation}
together with the techniques of Ref.~\cite{Scherer2004B} 
we can show that the modified  
assumption (a) necessarily implies that
\begin{eqnarray} \label{eq:specialloopcondition1}
&&\forall\, n\in\natuerl\:\:
\forall\,\mu_0,\mu',\mu''\quad\mbox{with}\quad
\mu'\not=\mu'':\;\;\\
&&\left|\mbox{Tr}\left[P_{\mu''}(U^nP_{\mu_0}U^{\dagger\,n})P_{\mu'}
(U^nP_{\mu_0} U^{\dagger\,n})P_{\mu''}\right]\right|<d\epsilon\:. \nonumber
\end{eqnarray}
This condition is equivalent to 
\begin{equation} \label{eq:specialloopcondition2}
\forall\, n\in\natuerl\:\:
\forall\,\mu_0,\mu',\mu''\quad\mbox{with}\quad
\mu'\not=\mu'':\;\;
\parallel P_{\mu'}(U^nP_{\mu_0}U^{\dagger\,n})P_{\mu''}\parallel_2
<\sqrt{d\,\epsilon}\:,
\end{equation}
where $\parallel A\parallel_2:=\sqrt{\mbox{Tr}[A^{\dagger}A]}$ denotes the 
Hilbert-Schmidt operator norm for any operator $A$. It then follows,  
{\em for all} $n\in\natuerl$ and {\em for all} $\mu_0,\mu''$, that
\begin{eqnarray}
\parallel\,[(U^nP_{\mu_0}U^{\dagger\,n}),P_{\mu''}]\,\parallel_2&=&
\parallel
(U^nP_{\mu_0}U^{\dagger\,n})P_{\mu''}-P_{\mu''}(U^nP_{\mu_0}U^{\dagger\,n})
\parallel_2\nonumber\\
&=&\parallel(\,\sum_{\mu'}P_{\mu'}\,)(U^nP_{\mu_0}U^{\dagger\,n})P_{\mu''}
-P_{\mu''}(U^nP_{\mu_0}U^{\dagger\,n})(\,\sum_{\mu'}P_{\mu'}\,)\parallel_2\nonumber\\
&\le&\sum_{\mu'\atop\mu'\not=\mu''}\parallel
P_{\mu'}(U^nP_{\mu_0}U^{\dagger\,n})P_{\mu''}-
P_{\mu''}(U^nP_{\mu_0}U^{\dagger\,n})P_{\mu'}\parallel_2\nonumber\\
&\le&\sum_{\mu'\atop\mu'\not=\mu''}\{\,\parallel
P_{\mu'}(U^nP_{\mu_0}U^{\dagger\,n})P_{\mu''}\parallel_2+\parallel
P_{\mu''}(U^nP_{\mu_0}U^{\dagger\,n})P_{\mu'}\parallel_2\}\nonumber\\
&<&2(\,\sharp\,\mu'\,)\sqrt{d\epsilon}\le 2d\sqrt{d\epsilon}=2d^{\frac{3}{2}}\sqrt{\epsilon}\;, 
\end{eqnarray}
i.e. $\,\parallel\,[(U^nP_{\mu_0}U^{\dagger\,n}),P_{\mu''}]\,\parallel_2<
2d^{\frac{3}{2}}\sqrt{\epsilon}\;\;$ {\em for all} $n\in\natuerl$ and 
{\em for all} $\mu_0,\mu''$. We thus get the following generalisation of 
the implication (a)$\Rightarrow $(b) of Theorem 1:

\noindent{\bf Theorem 2:}
{\it Let a projective partition}  $\{P_{\mu}\}$ 
{\it of a finite dimensional Hilbert space $\cal H$, a unitary 
map} $U$ {\it on} $\cal H$, {\it and a small $\epsilon>0$ be given. 
Then}
\begin{equation*}
\hspace{-4cm}
\forall \,\rho\in
\mathcal{S}_{\{P_{\mu}\}}\;
\forall\, k\in\hspace{-0.5mm}\natuerl \;\;\forall\,h_{\balphas},
h_{\bbetas}\in\mathcal{K}[\{P_{\mu}\}\,;\,k\,]\;\;\,\mbox{with}\;\;
h_{\balphas}\not=h_{\bbetas}\::
\end{equation*} 
\begin{equation}
\;\left|\mathcal{D}_{U,\rho}\,[h_{\balphas},h_{\bbetas}]\right|
<\epsilon\:\frac{\left(\mathcal{D}_{U,\rho}\,[h_{\balphas},h_{\balphas}]
\mathcal{D}_{U,\rho}\,[h_{\bbetas},h_{\bbetas}]\right)^{\frac{1}{2}}}
{|\mathcal{K}[\{P_{\mu}\}\,;\,k\,]\,|}
\end{equation} 
{\it only if}
\begin{equation}
\forall\,P_{\mu'}, P_{\mu''}\in\{P_{\mu}\}\;\,\forall\, n\in\natuerl
\,:\;\;\parallel\,[(U^nP_{\mu'}U^{\dagger\,n}),P_{\mu''}]\,\parallel_2
<2d^{\frac{3}{2}}\sqrt{\epsilon}\:,
\end{equation}
{\it where} $d=\mbox{dim}\,{\cal H}$ {\it and $\parallel\cdot\parallel_2$ denotes 
the Hilbert-Schmidt operator norm}.

\end{document}